\documentclass[12pt]{article}
\usepackage{amsmath}
\usepackage{amsfonts}
\usepackage{amssymb}
\usepackage{latexsym}
\usepackage{graphicx}
\usepackage[english]{babel}
\input epsf.sty

\begin{document}

\title{
{Hawking-Moss Tunneling in Noncommutative Eternal Inflation}}
 \vspace{3mm}
\author{{Yi-Fu Cai$^1$\footnote{caiyf@mail.ihep.ac.cn}, Yi Wang$^{2,3}$\footnote{wangyi@itp.ac.cn}}\\
{\small $^{1}$ Institute of High Energy Physics, Chinese Academy
of
Sciences, Beijing 100049, P. R. China}\\
{\small $^{2}$ Institute of Theoretical Physics, Chinese Academy
of
Sciences, Beijing 100080, P. R. China}\\
{\small $^{3}$  The Interdisciplinary Center for Theoretical Study of China (USTC), }\\
{\small Hefei, Anhui 230027, P. R. China}}
\date{}
\maketitle

\begin{abstract}

The quantum behavior of noncommutative eternal inflation is quite
different from the usual knowledge. Unlike the usual eternal
inflation, the quantum fluctuation of noncommutative eternal
inflation is suppressed by the Hubble parameter. Due to this, we
need to reconsider many conceptions of eternal inflation. In this
paper we study the Hawking-Moss tunneling in noncommutative
eternal inflation using the stochastic approach. We obtain a
brand-new form of the tunneling probability for this process and
find that the Hawking-Moss tunneling is more unlikely to take
place in the noncommutative case than in the usual one. We also
conclude that the lifetime of a metastable de-Sitter (dS) vacuum
in the noncommutative spacetime is longer than that in the
commutative case.

\end{abstract}

\maketitle

\section{Introduction}

Inflation has been widely considered as a remarkably successful
theory in explaining many problems in the very early universe,
such as the flatness, horizon and monopole problems
\cite{Guth81,Sato:1980yn,Linde82,Steinhardt82,Starobinsky-inf}.
During inflation, quantum effects play a crucial role and may
bring the universe into a self-reproducing process which is dubbed
``eternal inflation"
\cite{steinhardt-nuffield,vilenkin-eternal,linde-eternal}. The
string theory landscape indicates that there are a huge number of
meta-stable vacua surrounded by various kinds of effective
potentials
\cite{bousso-landscape,kachru-landscape,susskind-landscape,douglas-landscape}.
The realization of string landscape provides an important arena
for eternal inflation.

In eternal inflation driven by the false vacuum, the false vacuum
is a meta-stable state and would decay through a mix of
semiclassical tunneling and stochastic evolution. The probability
of finding the inflaton at the top of the plateau in its potential
decreases exponentially with time \cite{Guth:1985ya}. However, the
false vacuum is also expanding exponentially while decaying. When
the rate of exponential expansion is larger than the decay rate
during this process, the total volume of the false vacuum will
grow eternally although the false vacuum is decaying. In this
case, bubbles form by random nucleation and then start to expand.
Every growing bubble can be viewed as an open FRW universe
\cite{Coleman:1980aw}, and we are living in one of such ``pocket
universes" \cite{Guth:2000ka}. Another approach to eternal
inflation is achieved in chaotic inflation when the quantum
fluctuation of the inflaton dominates over its classical motion.
As the inflaton is rolling down the potential classically, its
change during one Hubble time ($\delta t=\frac{1}{H}$) can be
divided into $\delta\varphi=\Delta\varphi+\delta_q\varphi$, where
$\Delta\varphi$ denotes the classical value and $\delta_q\varphi$
represents the quantum one. For a Gaussian probability
distribution, when $\delta_q\varphi>0.61 \Delta\varphi$, the
quantum behavior overwhelms the classical evolution and inflation
becomes eternal.

When dealing with the decaying process of false vacua, one have to
make the inflaton tunnel from one false vacuum to another. One
method was provided by Coleman and De Luccia (CDL)
\cite{Coleman:1980aw}; another method was investigated by Hawking
and Moss \cite{Hawking:1981fz}. During the Hawking-Moss tunneling,
the potential between two vacua is so flat that CDL instanton can
not exist. It was shown that the probability of tunneling from one
false vacuum $\varphi_0$ to another is given by
\begin{eqnarray}\label{Pclass}
P_C &\sim&
\exp\left(-\frac{24\pi^2}{V(\varphi_0)}+\frac{24\pi^2}{V(\varphi_{\rm top})}\right)\nonumber\\
&\sim& \exp\left(-8\pi^2\cdot\frac{H(\varphi_{\rm
top})^2-H(\varphi_0)^2}{H(\varphi_0)^2H(\varphi_{\rm
top})^2}\right)~,
\end{eqnarray}
which is related to the value of the top barrier. A proper
scenario of this tunneling can be realized in the stochastic
approach to inflation
\cite{Starobinsky:1986fx,Rey:1986zk,Nakao:1988yi,Linde:1991sk,Gratton:2005bi,Linde:2006nw}.
Here the potential of the scalar field is flat enough for slow
rolling which requires $V''<V$. In the stochastic description, the
quantum fluctuations can be simulated by the stochastic noise and
the scalar field walks in random.

As is known, eternal inflation happens when the energy scale of
the universe is extremely high. Thus we would like to take into
consideration more fundamental theories in logic, namely, the
string theory. Recently, we have considered the effects of
spacetime noncommutativity on slow-roll eternal inflation in Ref.
\cite{Cai:2007et}. In noncommutative inflation, it is generally
assumed that the background evolution of inflaton is not modified
but the fluctuations are affected by noncommutativity (see Ref.
\cite{Bran2002,Koh:2007rx,qghuang-nc,cai2007}, for a review in
\cite{Brandenberger:2007rg} and references therein). In order to
introduce the noncommutativity \cite{Li:1996rp,Seiberg99} into the
4-dimensional flat Friedmann-Robertson-Walker universe, we would
like to define another time coordinate $\tau$,
\begin{equation}
  ds^2=dt^2-a^2(t)d{\vec x}^2=a^{-2}(\tau)d\tau^2-a^2(\tau)d{\vec x}^2~,
\end{equation}
where $a$ is the scale factor. Then the spacetime uncertainty
relation can be realized by the commutation relation:
\begin{equation}
  [\tau,x]_*=i M_{\rm N}^{-2}~,
\end{equation}
where $M_{\rm N}$ is the energy scale of noncommutativity and the
$*$-product is defined as
\begin{eqnarray}
  (f*g)(x,\tau) &=& \exp \left(-\frac{i}{2}M_{\rm N}^{-2}\left(\partial_x\partial_{\tau'}
  -\partial_\tau\partial_y\right)\right) \nonumber\\
  &\times& f(x,\tau)g(y,\tau') \left. \right|_{y=x,\tau'=\tau}~.
\end{eqnarray}
From the result of the paper \cite{Cai:2007et}, we can see that
the quantum fluctuation still satisfies the Gaussian distribution,
but the form of its amplitude in IR region is changed to
$\delta_q\varphi\simeq\frac{1}{2\pi}\frac{M_{\rm N}^2}{H}$.
Therefore, when the Hubble parameter is lifted highly enough,
eternal inflation would cease. This is strongly different from the
normal scenario of eternal inflation. In this paper, we shall use
the analysis of noncommutativity mentioned above (especially the
IR region) and the stochastic approach to study Hawking-Moss
tunneling of eternal inflation.

This paper is organized as follows. In Section 2, we study the
Hawking-Moss tunneling in noncommutative eternal inflation using
the stochastic approach. In Section 3, we make conclusions that
Hawking-Moss tunneling is more unlikely to happen in the
noncommutative case than in the usual one and the lifetime of a
metastable de-Sitter vacuum in the noncommutative spacetime is
longer than that in the commutative case.

\section{Stochastic Approach to Noncommutative Inflation}

One can describe inflation by analyzing the stochastic probability
distribution $P(\varphi,t)$, which represents the probability to
find the inflaton field $\varphi$ at the time $t$. In our note we
consider the probability distribution averaged in a Hubble volume
observed by a comoving observer. The inflaton evolves as a
Brownian particle. Consequently, the probability distribution
$P(\varphi,t)$ satisfies the Fokker-Planck (FP) equation(see
detailed introduction in Ref. \cite{Linde:1991sk,fpbrownian}):
\begin{eqnarray}\label{FPeq}
\frac{\partial P}{\partial
t}=\frac{\partial}{\partial\varphi}\left(
\frac{\partial(DP)}{\partial\varphi}+\gamma\frac{dV}{d\varphi}P
\right)~,
\end{eqnarray}
where $D$ is the diffusion coefficient and $\gamma$ is the
mobility coefficient\footnote{A general form of FP equation is
given by
\begin{eqnarray}
\frac{\partial P}{\partial
t}=\frac{\partial}{\partial\varphi}\left(D^{1-d}
\frac{\partial(D^dP)}{\partial\varphi}+\gamma\frac{dV}{d\varphi}P
\right)~.\nonumber
\end{eqnarray}
However, the choice of the parameter $d$ does not affect the
calculation of probability distribution a lot. Therefore we do not
plan to discuss it in our note but only focus on the case $d=1$.}.
Using slow roll approximation $\dot\varphi\simeq -V'/(3H)$, we can
establish that $\gamma=\frac{1}{3H}$. In the following we need to
derive the form of the coefficient $D$.

Following the usual knowledge of inflation, the background
evolution of noncommutative inflation can be described by
\begin{eqnarray}
  3H^2\simeq V(\varphi)\ ,\label{f}
\end{eqnarray}
where we take the normalization $M_p^2=1/8\pi G=1$. According to
the calculation in Ref. \cite{Cai:2007et}, the IR quantum
fluctuation in the momentum space $\delta_q\varphi_k$ is linked to
the canonical perturbation $u_k$ by $u_k\simeq
a\delta_q\varphi_k$, and when the perturbation begins to be
generated the initial conditions require $u_k$ to be canonically
normalized as $u_k\simeq \frac{1}{\sqrt{2k}}$ with $a\simeq
Hk/M_{\rm N}^2$. Therefore, the IR quantum fluctuation in momentum
space can be generally given by,
\begin{eqnarray}\label{deltaqm}
\delta_q\varphi_k\simeq \frac{1}{\sqrt{2k}}\frac{M_{\rm
N}^2}{Hk}~.
\end{eqnarray}
After that, the fluctuations outside the horizon are nearly frozen.
It can be shown that the initial wave length for the $k$ mode is
$\lambda_k=H/M_{\rm N}^2$ \cite{Cai:2007et}, so it is appropriate to
do the spacial average at a length scale $H/M_{\rm N}^2$. During one
Hubble time, we can calculate the IR quantum fluctuation in
coordinate space $\delta_q\varphi$ as follows,
\begin{eqnarray}\label{deltaqH}
\delta_q\varphi|_H&\equiv&\sqrt{\left<\delta_q\varphi^2\right>}|_H=\left(\int_{k=aM_{\rm
N}^2/H}^{k=e\times aM_{\rm N}^2/H}
\frac{dk}{k}\frac{k^3}{2\pi^2}\delta_q\varphi_k\delta_q\varphi_{-k}\right)^{\frac{1}{2}}\nonumber\\&\simeq&\frac{1}{2\pi}\frac{M_{\rm
N}^2}{H}~.
\end{eqnarray}

Note that due to the nearly scale invariance of the spectrum, the
result of (\ref{deltaqH}) is not sensitive to the length scale
$H/M_{\rm N}^2$ where we do the spacial average. We might have
chosen the Hubble length $H^{-1}$ or the noncommutativity scale
$M_{\rm N}^{-1}$, and the result of the integration in leading
order does not change.

Besides, since $M_{\rm N}^{-1}$ arises as another important time
scale rather than $H^{-1}$, it is worthy to calculate the quantum
fluctuations during the time $M_{\rm N}^{-1}$. We have
\begin{eqnarray}\label{deltaqn}
\delta_q\varphi|_{M_{\rm
N}}&\equiv&\sqrt{\left<\delta_q\varphi^2\right>}|_{M_{\rm
N}}=\left(\int_{k=aM_{\rm N}^2/H}^{k=e^{H/M_{\rm N}}\times aM_{\rm
N}^2/H}
\frac{dk}{k}\frac{k^3}{2\pi^2}\delta_q\varphi_k\delta_q\varphi_{-k}\right)^{\frac{1}{2}}\nonumber\\&\simeq&\frac{1}{2\pi}\frac{M_{\rm
N}^{\frac{3}{2}}}{H^{\frac{1}{2}}}~,
\end{eqnarray}
which is a bit different from Eq. (\ref{deltaqH}). We will show
later, however, this is also insensitive to the simulation of
Langevin equation.

We can simulate the quantum fluctuation in IR region by the
Langevin equation which is expressed as
\begin{eqnarray}\label{m}
\dot\varphi \simeq -\frac{V'}{3H}-\frac{M_{\rm N}^2}{3
H^{\frac{1}{2}}}\eta~.
\end{eqnarray}
Here $\eta$ is a stochastic noise term added to simulate the quantum
fluctuation of the inflaton and we make its form to be Gaussian
satisfying
\begin{eqnarray}
<\eta(t)>=0~,~~~<\eta(t)\eta(t')>=\frac{9}{4\pi^2}\delta(t-t')~.
\end{eqnarray}
This simulation is quite general and very efficient in the IR region
of noncommutative inflation no matter what time scale we use. When
we consider the quantum fluctuations in one Hubble scale, we can
recover $<\delta_q\varphi^2>\simeq M_{\rm N}^4/(4\pi^2 H^2)$ with
the fluctuation of $\varphi$ integrated in one Hubble time,
\begin{equation}
  \delta_q\varphi = -H\int^{1\over H} \frac{M_{\rm N}^2\eta}{3H^{\frac{3}{2}}}dt\
  ;
\end{equation}
moreover, when we consider the quantum fluctuations in one
noncommutative scale, then we can recover $<\delta_q\varphi^2>\simeq
M_{\rm N}^3/(4\pi^2 H)$ with the fluctuation of $\varphi$ integrated
during the time scale $1/M_{\rm N}$,
\begin{equation}
  \delta_q\varphi = -M_{\rm N}\int^{1\over M_{\rm N}} \frac{M_{\rm N}\eta}{3H^{\frac{1}{2}}}dt\
  .
\end{equation}

Therefore, we have the expression $\frac{d}{dt}<\varphi^2>\simeq
M_{\rm N}^4/(4\pi^2 H)$ both in the Hubble scale and the
noncommutative scale. According to the property of Brownian
motion, the diffusion coefficient $D$ is approximately equal to
half of $\frac{d}{dt}<\varphi^2>$, and thus we have $D=M_{\rm
N}^4/(8\pi^2 H)$.

For simplicity, we study the stationary ansatz of Eq.
(\ref{FPeq}): $\partial_t P_{\rm N}=0$. Consequently the FP
equation (\ref{FPeq}) in IR region of noncommutative case can be
solved as
\begin{eqnarray}
P_{\rm N}(\varphi,t)&\sim& \exp\left\{-\frac{8\pi^2}{3M_{\rm N}^4}
(V-V_0)
%+
%\frac{1}{2}\ln V
\right\}\nonumber\\
&\sim&\exp\left\{-8\pi^2\cdot\frac{H^2-H_0^2}{M_{\rm
N}^4}\right\}~,\label{FPsol}
\end{eqnarray}
where $V_0$ (and $H_0$) appears from a proper normalization.

Keeping in mind that $H>M_{\rm N}$ in the IR region of
noncommutative eternal inflation, and comparing the denominator on
the exponential of the noncommutative result (\ref{FPsol}) with
the commutative result (\ref{Pclass}), we conclude that the
tunneling probability is more suppressed by the spacetime
noncommutativity. This suppression has clear physical
interpretation. Since the spacetime noncommutativity generally
suppresses the quantum fluctuation of the inflaton, it should make
quantum behaviors of the inflaton, such as the Hawking-Moss
tunneling, more unlikely to happen.

\begin{figure}[htbp]
\includegraphics[width=3.7in]{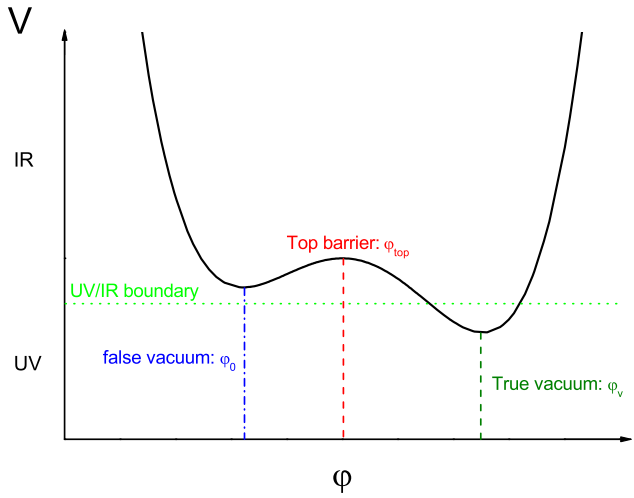}
\caption{A sketch map of Hawking-Moss tunneling from a false
vacuum $\varphi_0$ to the true one $\varphi_v$. When the inflaton
$\varphi$ lies above the green dot line, its distribution function
satisfies Eq. (\ref{FPsol}) and the probability of tunneling is
suppressed exponentially with respect to $V$; meanwhile, if
$\varphi$ is placed below the green line, the form of distribution
function returns the usual one (\ref{Pclass}).} \label{fig:tunnel}
\end{figure}

We also note that the equation (\ref{Pclass}) and (\ref{FPsol})
can be linked smoothly describing the energy density crossing the
noncommutative UV/IR boundary. It is known that when $H<M_{\rm
N}$, the noncommutative inflation is in the UV region, and it is
in the IR region when $H>M_{\rm N}$. The probability distribution
of inflaton in the UV region is described by Eq. (\ref{Pclass}) in
which the maximal value of the potential is $V\simeq 3H^2=3M_{\rm
N}^2$. This is just the minimal value of the potential in the IR
region. Consequently the distribution function of noncommutative
eternal inflation in the whole parameter space is continuous, and
hence, there is no pathology when the field $\varphi$ tunnels
through the UV/IR boundary(See Fig. \ref{fig:tunnel}).

In order to make this result more explicitly and to investigate the
details of physics, we consider the example
$V(\varphi)=\lambda\varphi^4$ when $\varphi$ is near one minimal of
the potential. To solve the equations (\ref{f}) and (\ref{m}), we
define for simplicity $\sigma\equiv\varphi^2$, then we have
\begin{equation}
  \dot\sigma+\alpha\sigma+\beta\eta=0\ ,\ \ \alpha\equiv
  8\sqrt\frac{\lambda}{3}\ ,\ \
  \beta\equiv\frac{2}{3}M_{\rm
N}^2\sqrt[4]{\frac{3}{\lambda}}\
  .\label{seq}
\end{equation}
The solution of (\ref{seq}) can be written as
\begin{equation}
  \sigma(t)=\sigma_0 e^{-\alpha t}+\beta e^{-\alpha t}\int_0^t e^{\alpha
  t_1}\eta(t_1)dt_1\ .
\end{equation}
where $\sigma_0\equiv\sigma(0)$ sets the initial condition at $t=0$.

Following \cite{Li:2007uc}, the distribution function for $\sigma$
can also be given by
\begin{eqnarray}
  P_{\rm N}(\sigma_m) &\sim& \int [d\eta]dt \exp\left(-\frac{2}{9}\pi^2\int_0^\infty
  dt_1\eta^2(t_1)\right) \nonumber\\
  &\times& \delta\left(\sigma(t)-\sigma_m\right)\ ,
\end{eqnarray}
which denotes the number of times the universe arrives at the
$\sigma(t) = \sigma_m$ surface during infinite time. By using
$\delta(y)=\int\frac{dx}{2\pi}e^{ixy}$, and doing Gaussian
integration twice, we obtain the integral
\begin{eqnarray}
  P_{\rm N}(\sigma_m) &\sim& \int dt ~ \sqrt{\frac{8\pi\lambda}{3M_{\rm
N}^4(1-e^{-2\alpha
  t})}} \nonumber\\
  &\times& \exp\left(-\frac{8\pi^2\lambda}{3 M_{\rm
N}^{4}}\sigma_m^2\frac{\left(e^{\alpha
t}-\frac{\sigma_0}{\sigma_m}\right)^2}{e^{2\alpha
  t}-1}\right) .
\end{eqnarray}

This integral seems problematic because there is a divergence when
$t\rightarrow\infty$. However, note that the energy scale of
eternal inflation along a comoving world line will eventually
drop. As $t$ becomes larger, inflation enters the UV region, and
the behavior of evolution returns to the commutative case.
Consequently, the full integral does not suffer from the
divergence, so this measure is well defined.

Further, what we care about is the tunneling probability which
corresponds to the case: $\sigma_m>\sigma_0$. By using the saddle
point approximation on the exponential as in \cite{Li:2007uc}, we
obtain
\begin{eqnarray}
  P_{\rm N}(\sigma_0)&\sim&
  %\frac{1}{\alpha\sigma_0}
  \exp\left(-8\pi^2 M_{\rm
N}^{-4}(H_m^2-H_0^2)\right)\nonumber\\
  &\simeq&\exp\left\{-\frac{8\pi^2}{3M_{\rm N}^4}
  (V_m-V_0)\right\}\ ,
\end{eqnarray}
which is consistent with Eq. (\ref{FPsol}).

\section{Conclusion and Discussions}

To take a further discussion, we would like to compare the
difference of dS decaying processes whether or not the spacetime
noncommutativity is present. According to the work of Coleman and De
Luccia\cite{Coleman:1980aw}, the decay time of a metastable dS
vacuum has an approximate expression $T\sim P^{-1}$. By neglecting
all the sub-exponential factors, we have
\begin{eqnarray}
T_C&=&\exp\left\{24\pi^2M_p^4(\frac{1}{V_0}-\frac{1}{V_{top}})\right\};\\
\label{timenc}T_{\rm N}&=&\exp\left\{\frac{8\pi^2}{3M_{\rm N}^4}
(V_{top}-V_0)\right\}~,
\end{eqnarray}
which represent the dS decay times without the spacetime
noncommutativity, and the IR region with noncommutativity
respectively. To be clear, we have written $M_p$ explicitly here.
It is clear that $T_N>T_C$, which is a very general result
indicating that the lifetime of a metastable dS vacuum with
noncommutativity is longer than that without noncommutativity.

To summarize, from the results obtained in this note we learn that
the Hawking-Moss tunneling effect of noncommutative eternal
inflation in the IR region is greatly different from the usual
one. Its probability distribution is exponentially suppressed by
the top barrier value of the potential and make Hawking-Moss
tunneling more difficult to happen than in the usual case. This is
because the quantum fluctuation is suppressed by spacetime
noncommutativity. Consequently, we may expect the application of
noncommutativity would bring a closer sight into high energy
physics of eternal inflation. Based on the new form of the
probability distribution, we find that the lifetime of a
metastable dS vacuum in the noncommutative case is longer than in
the usual one. This may leave more clues for investigating the new
physics of noncommutativity which is worthy for further studies.

\section*{Acknowledgments}

We thank Robert Brandenberger, Yun-Song Piao, Miao Li and Xinmin
Zhang for valuable comments. We also thank the anonymous referee
for valuable suggestions on our manuscript. This work is supported
in part by National Natural Science Foundation of China under
Grant Nos. 90303004, 10533010, 19925523 and 10405029, and by the
Chinese Academy of Science under Grant No. KJCX3-SYW-N2. The
author Y.W. acknowledgments grants of NSFC.

%\vfill

\end{document}